\documentstyle[multicol,epsf,aps]{revtex}
\begin{document}
\draft
\title{Halflife of  $^{56}$Ni in cosmic rays}

\author{J. Lund Fisker, G. Mart\'{\i}nez-Pinedo and K. Langanke}

\address{Institute of Physics and Astronomy, University of {\AA}rhus,
  Denmark}

\date{\today} 

\maketitle

\begin{abstract}
  A measurement of the $^{56}$Ni cosmic ray abundance has been
  discussed as a possible tool to determine the acceleration time
  scale of relativistic particles in cosmic rays. This conjecture will
  depend on the halflife of totally ionized $^{56}$Ni which can only
  decay by higher-order forbidden transitions. We have calculated this
  halflife within large-scale shell model calculations and find $t_{1/2}
  \approx 4 \times 10^4$ years, only slightly larger than the currently
  available experimental lower limit, but too short for $^{56}$Ni to
  serve as a cosmic ray chronometer.
\end{abstract}


\begin{multicols}{2}
  
\begin{figure}
  \begin{center}
    \leavevmode
    \epsfxsize=0.6\columnwidth
    \epsffile{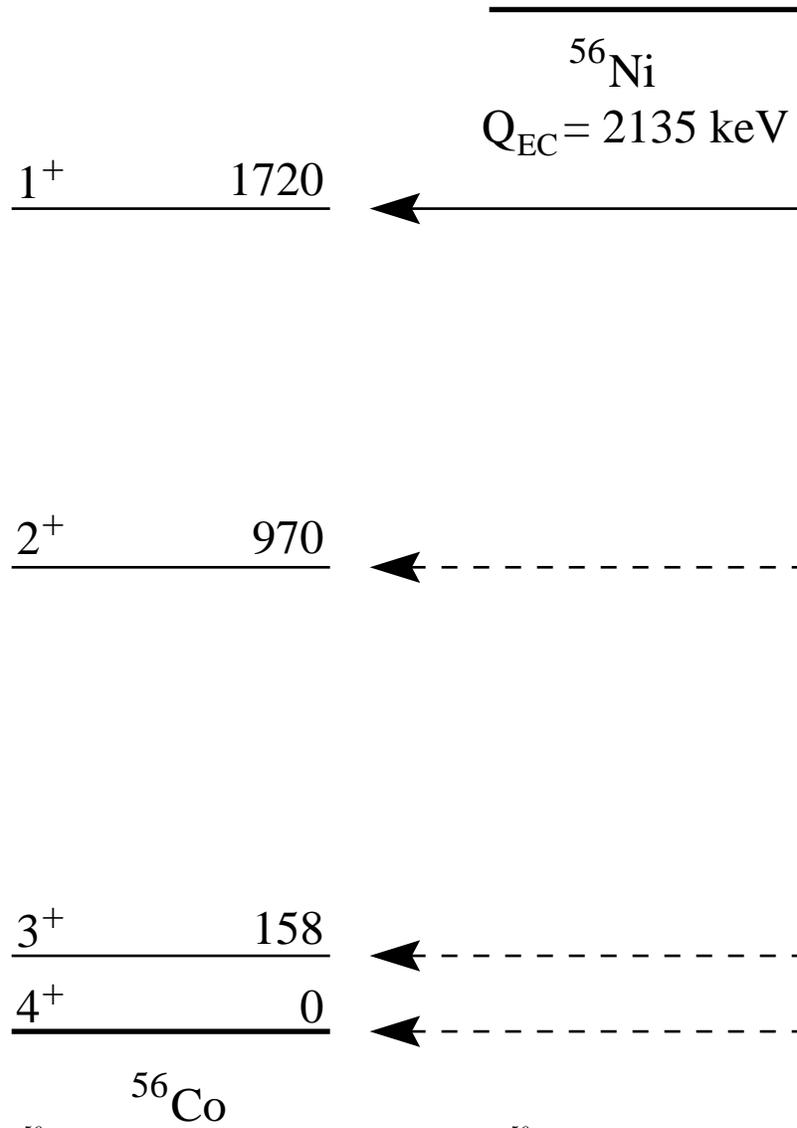}
    \caption{Level scheme of $^{56}$Co relevant for the decay of totally
      ionized $^{56}$Ni. The solid line indicates the decay of $^{56}$Ni
      under laboratory conditions, while the dashed lines show the decay
      branches for fully ionized $^{56}$Ni.}
    \label{fig1}
  \end{center}
\end{figure}

In the laboratory $^{56}$Ni decays by electron capture to the $1^+$
state in $^{56}$Co at 1.72 MeV ($\approx 100 \%$ branch) with a
halflife of $t_{1/2}=6.075\pm0.020$ days \cite{ni56t}.  If $^{56}$Ni
is, however, stripped of its electrons, this transition is no longer
energetically allowed (see Fig.~\ref{fig1}), and the decay is assumed
to occur to the $3^+$ state at 158 keV via a second forbidden unique
transition. Currently only a lower limit for the halflife of totally
ionized $^{56}$Ni could be established ($2.9 \times 10^4$ y)
\cite{Norman90}, but from systematics of similar decays in other
nuclei it has been argued that the halflife can be as large as $5.4
\times 10^6$ y \cite{Norman90}.

The drastic change in halflife, if stripped of electrons, makes
$^{56}$Ni a potential chronometer for cosmic rays. It is generally
believed that supernovae are the site for the acceleration of
relativistic particles found in cosmic rays; during this acceleration
the nuclei are stripped of electrons. $^{56}$Ni is very abundantly
produced in supernovae and its decay (with $t_{1/2}=6$ days) powers
the early supernova lightcurve. However, if supernovae are the site
for the acceleration of relativistic cosmic-ray particles and if the
associated acceleration timescale is short enough, some of the
$^{56}$Ni will survive in cosmic rays (now being totally ionized and
hence having a significantly larger halflife), thus making a future
measurement of the $^{56}$Ni abundance in cosmic rays a potential
chronometer for this process. Nevertheless, the $^{56}$Ni cosmic ray
abundance will also depend on the survival rate of the nucleus after
acceleration.  As the time the cosmic rays stay in our galaxy is in
the range 10-20~Myr~\cite{mn54exp}, a halflife of totally ionized
$^{56}$Ni shorter than this timescale would lead to a further
depletion of the $^{56}$Ni cosmic ray abundance.

It is the aim of this note to calculate the halflife of totally
ionized $^{56}$Ni. As weak interaction transitions in general are
sensitive to nuclear correlations, our method of choice is the
interacting shell model.  Shell model diagonalization approaches have
made significant progress in recent years and, combined with improved
computer technologies, basically allow now for complete $pf$ shell
calculations in the mass $A=56$ mass range, e.g. \cite{Mn54}.  For
example, using the KB3 interaction~\cite{Zuker} one calculates the
(laboratory) halflife of $^{56}$Ni against electron capture as
6.7~d~\cite{Langanke98a}, which is in good agreement with experiment.

As the theory for nuclear beta decay is well developed and the
respective formulae for allowed and forbidden transitions can be found
in many textbooks, e.g. \cite{Behrens}, it is not necessary to repeat
it here. Just short of a complete calculation in the $pf$ shell we
have performed large-scale shell model calculations for the $^{56}$Ni
ground state and the 3 lowest states in $^{56}$Co in a model space
which allowed a maximum of 6 particles to be excited from the
energetically favored $f_{7/2}$ orbit to the rest of the $pf$ shell in
the final nucleus.  We expect that the transition matrix elements are
converged at this level of truncation (for an example see Fig. 2 of
Ref.~\cite{Mn54}).  The decay to the $3^+$ state in $^{56}$Co is a
unique second-forbidden transition, while those to the $4^+$ ground
state and the $2^+$ state at 970 keV are non-unique fourth-forbidden
and non-unique second-forbidden transitions \cite{Behrens},
respectively.  Shell model codes solve the many-body problem in Fock
space. Thus the calculation of the transition matrix elements requires
assumptions about the single particle wave functions. Here we have
assumed harmonic oscillator wave functions with an oscillator
parameter, $b=1.99$~fm, determined from the $^{56}$Ni charge radius
\cite{deVries} using the prescription of Ref.  \cite{Towner}; but we
have also considered Woods-Saxon radial wave functions derived from a
potential which includes spin-orbit and Coulomb terms \cite{Bohr}.
For the Gamow-Teller strength to be reproduced in complete
$0\hbar\omega$ calculations it is necessary to renormalize the spin
operators by a universal factor $(0.74)^2$
\cite{Wildenthal,Langanke95,Martinez96}.  But such a renormalization
has not been established for forbidden transitions and we thus assume
no quenching.

We have performed the shell model calculations with 
two different residual interactions. At first we used the KB3 interaction,
but we also performed calculations with a new version of this force in
which some slight monopole deficiencies have been corrected
\cite{Caurier99}. Our calculations reproduce the $^{56}$Co level scheme
rather well as we place the excited $2^+$ and $3^+$
states at 216 (237) keV and 1.03 (1.06) MeV, respectively (the energies
in paranthesis refer to the KB3 interaction).
In the calculation of the various partial halflives for totally ionized
$^{56}$Ni we, however, used the experimental energies.
Our results for the various beta decays are summarized in
Table~\ref{tab:tab1}. 

As expected \cite{Norman90} totally ionized $^{56}$Ni decays
preferably to the $3^+$ state at 158 keV. The other possible
second-forbidden transition is disfavored by phase space due to the
significantly smaller Q-value, while the transition to the ground
state is strongly suppressed as it is of fourth order. For the
dominating decay all four shell model calculations are in  close
agreement. We just find a halflife of totally ionized $^{56}$Ni of $4
\times 10^4$ y, which is only slightly larger than the current lower
limit ($2.9 \times 10^4$ y, \cite{Norman90}) and thus might be in
experimental reach.  

We like to add two short remarks concerning the other transitions.
First, we note that all formfactors of the type {}$^VF_{KK-11} $ (in
the notation of~\cite{Behrens}) are identical zero for harmonic
oscillator wave functions, but not for Woods-Saxon single particle
states.  In our case this applies to the {}$^VF_{211}$ and
{}$^VF_{431}$ in the calculations of the non-unique second and fourth
transitions respectively. It turns out that the contributions of these
formfactors are not negligible and cause the differences observed in
the partial halflives of these two transitions calculated with
Woods-Saxon and harmonic oscillator wave functions.  Second, the
differences observed between the two interactions in the decay to the
ground state and the $2^+$ excited state are due to cancellations in
the calculations of the ${}^AF_{KK1}$-type matrix elements using the
KB3 interaction. Such a cancellation is absent for the modified
interaction.  Such a case does not occur for the dominating decay to
the $3^+$ state and thus the shell model halflife should be rather
reliable.

Can $^{56}$Ni serve as a chronometer for cosmic rays? The answer
implied from our study is unfortunately negative, as the calculated
halflife is significantly shorter than the time relativistic particles
need to escape from our galaxy ($>10^7$ y).  Thus, even $^{56}$Ni,
originally accelerated from a supernova remnant into the cosmic rays,
will be depleted too fast to serve as cosmic ray chronometer. This
conclusion is not expected to change, even if forbidden transitions
require a renormalization of the spin operator. If we assume the same
universal renormalization factor like for Gamow-Teller transitions,
the lifetime of totally ionized $^{56}$Ni increases to $7.3 \cdot
10^4$ y, but is still much too small for $^{56}$Ni to substantially
survive in cosmic rays.

\acknowledgements

This work was supported in part by the Danish Research Council.

\end{multicols}

\begin{table}
  \begin{center}
    \caption{Partial halflives (in years) for fully stripped
      $^{56}$Ni. The calculations have been performed with the KB3
      interaction and its recently modified version
      \protect\cite{Caurier99} and employing harmonic oscillator
      (harm. osc.) and Woods-Saxon (WS) radial wave functions.}
    \label{tab:tab1}
    \renewcommand{\arraystretch}{1.1}
    \begin{tabular}{ccccc}
      final state & \multicolumn{2}{c}{HO} & \multicolumn{2}{c}{WS} \\
       \cline{2-3}\cline{4-5}
      & KB3 & mod. KB3 & KB3 &  mod. KB3 \\
      \hline
      $4^+$ & $2.6 \times 10^{12}$ & $3.2 \times 10^{12}$ & $4.1 \times
       10^{12}$ & $5.0 \times 10^{12}$ \\
      $3^+$ & $3.8 \times 10^4$ & $3.7 \times 10^4$ & $4.2 \times
       10^4$ & $3.9 \times 10^4$ \\
      $2^+$ & $5.6 \times 10^7$ & $1.1 \times 10^8$ & $1.2 \times
       10^8$ & $3.3 \times 10^8$ \\
    \end{tabular}
  \end{center}
\end{table}

\end{document}